\documentclass[12pt]{iopart}

\usepackage{iopams}  
\usepackage{color}
\usepackage[dvipsnames]{xcolor} 
\usepackage{tikz}
\usetikzlibrary{decorations.pathmorphing,patterns} 
\usepackage{psfrag}
\usepackage{graphicx}
\usepackage{stackrel}
\usepackage{amsfonts}
\usepackage{bm}          
\usepackage{amssymb}
\usepackage{perpage}
\MakePerPage{footnote}

\newcommand{\be}{\begin{equation}}
\newcommand{\ee}{\end{equation}}
\newcommand{\ba}{\begin{eqnarray}}
\newcommand{\ea}{\end{eqnarray}}

\begin{document}

\title[Potts-model critical manifolds revisited]
{Potts-model critical manifolds revisited}

\author{Christian R.\ Scullard$^{1}$ and Jesper Lykke Jacobsen$^{2,3}$}
\address{${}^1$Lawrence Livermore National Laboratory, Livermore CA 94550, USA}
\address{${}^2$LPTENS, \'Ecole Normale Sup\'erieure -- PSL Research University, 24 rue Lhomond, F-75231
Paris Cedex 05, France}
\address{${}^3$Sorbonne Universit\'es, UPMC Universit\'e Paris 6, CNRS UMR 8549, F-75005 Paris, France} 

\eads{\mailto{scullard1@llnl.gov}, \mailto{jesper.jacobsen@ens.fr}}

\begin{abstract}

We compute the critical polymials for the $q$-state Potts model on all Archimedean lattices,
using a parallel implementation of the algorithm of Ref.~\cite{Jacobsen14}
that gives us access to larger sizes than previously possible. 
The exact polynomials are computed for bases of size $6 \times 6$ unit cells,
and the root in the temperature variable $v={\rm e}^K-1$ is determined numerically
at $q=1$ for bases of size $8 \times 8$.
This leads to improved results for bond percolation thresholds, and for the
Potts-model critical manifolds in the real $(q,v)$ plane.
In the two most favourable cases, we find now the kagome-lattice threshold to eleven
digits and that of the $(3,12^2)$ lattice to thirteen. Our critical manifolds reveal many
interesting features in the antiferromagnetic region of the Potts model, and determine
accurately the extent of the Berker-Kadanoff phase for the lattices studied.

\end{abstract}

\noindent

\section{Introduction}
The computation of critical thresholds in percolation and the Potts model has long been the domain of Monte Carlo methods \cite{Lee08,SudingZiff99,ZiffSuding97} or transfer matrix techniques \cite{FengDengBlote08,Ding10} of similar accuracy. Recently, we developed a radically different technique called the method of critical polynomials \cite{ScullardZiff08,ScullardZiff10,Scullard11-2,Scullard11,Jacobsen12,ScullardJacobsen2012,JacobsenScullard2013} which gives us the ability to compute percolation and Potts-model thresholds to precisions that are orders of magnitude greater than that obtained with traditional tools. The idea is based on the conjecture that the roots of a particular graph polynomial, to be defined below, provide estimates for critical points that become more accurate by increasing the size of the graph on which it is computed. This conjecture is essentially equivalent \cite{ScullardJacobsen2012} with the universality conjecture for crossing probabilities \cite{Cardy92}, so it is probably not in serious doubt (at least for percolation). However, computing the polynomial on large graphs is a significant challenge for which, at present, the transfer matrix appears to be the best tool.

A recent advance in this area was made in \cite{Jacobsen14}, which allowed the polynomial to be calculated on graphs of hundreds of edges in some cases, producing estimates of Potts and percolation thresholds far exceeding that possible with traditional techniques, such as Monte Carlo. It also allows us to study the $v<-1$ regime of the Potts model on various lattices, something that was very difficult previously. Here we implement the algorithm of \cite{Jacobsen14} in parallel on the Cab and Vulcan supercomputers at Lawrence Livermore National Laboratory, improving the resolution of the phase diagrams and the accuracy of the percolation thresholds found in that work. Thus, this paper should be considered an addendum to \cite{Jacobsen14} and, although we will give a brief description of the critical polynomial method, the reader interested in learning the details of the transfer matrix algorithm is referred there.

\section{Critical polynomials}
We consider here the Potts model \cite{Potts52} in the Fortuin-Kasteleyn (FK) cluster representation \cite{FK1972}. On a graph $G=(V,E)$ composed of vertices $V$ and edges $E$, the partition function is given by
\begin{equation}
 Z=\sum_{A \subseteq E} v^{|A|} q^{k(A)} \,, \label{eq:Z}
\end{equation}
where the sum is over all the subsets, $A$, of $E$, $|A|$ is the number of edges in $A$ and $k(A)$ is the number of connected components including isolated vertices. Each term in the sum can be thought of as a configuration in which each edge contributes a factor of $v$ if it is present in the configuration (or ``open'') and a factor of $1$ otherwise. The sum (\ref{eq:Z}) restricted to configurations consistent with the occurence of a particular event is the {\it weight} of that event. On a finite graph, weights are polynomials in $q$ and $v$. 

\begin{figure}
\begin{center}
\includegraphics[width=12cm]{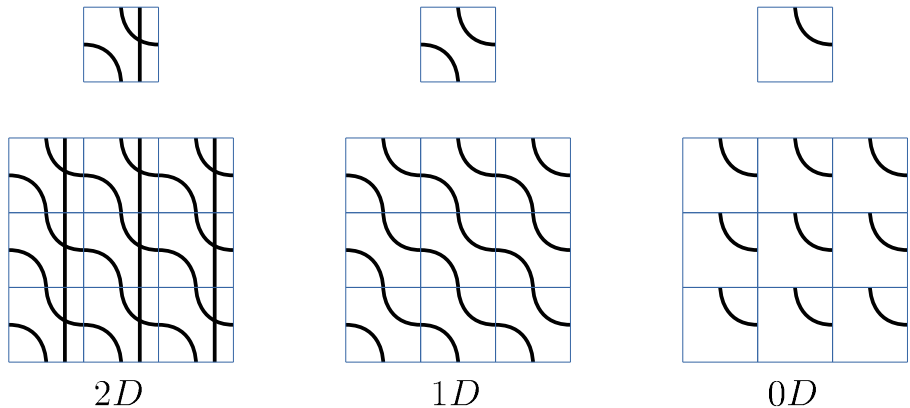}
\caption{The three possible global connectivity states that can arise from tiling a basis (shown on top) with a particular configuration.}
\label{fig:2D0D}
\end{center}
\end{figure}

To compute the critical polynomial on a given lattice, we must first choose a basis, $B$. This is a subgraph of the lattice along with a specification of how copies of $B$ are arranged to form the infinite lattice. This is defined by identifications between the vertices on the boundary of $B$, which we call terminals. Once $B$ is chosen, we compute the weights of two events; 2D and 0D. To understand what these are, consider an arbitrary configuration on $B$. If, after $B$ has been tiled to form the infinite lattice with each copy of $B$ containing the same configuration, every copy of $B$ is connected by open edges, we say that the event 2D occurs. Likewise, if we can traverse an infinite number of copies of $B$ but some remain unconnected, then we have the event 1D. Finally, if one cannot get anywhere on open edges, we have 0D. The weights of these events are denoted $P_{2D}(q,v)$, $P_{1D}(q,v)$ and $P_{0D}(q,v)$. These three possibilities are shown in Figure \ref{fig:2D0D}. The critical polynomial is then given by
\begin{equation}
 P_B(q,v)=P_{2D}(q,v)-q P_{0D}(q,v) .
\end{equation}

The root of this polynomial provides the estimates for the critical point. In the case of a lattice on which the critical point is known exactly \cite{Scullard06,Ziff06,ZiffScullard06}, the root of the polynomial provides the exact answer. Indeed, this method was motivated by the observation that all known exact solutions are roots of polynomials, and our graph polynomial always factorises in such cases, for any size of $B$, shedding a factor which corresponds to the polynomial of the exact solution.

On the other hand, for unsolved problems, the roots of the critical polynomial are approximations that become more accurate with the size of $B$. The estimates are generally very good, even for bases of moderate size, and they get better as we increase the number of edges in $B$. Using an extrapolation technique allows us to locate critical points with unprecedented accuracy. Although the fact that the roots of the polynomials should approach the exact critical point as the basis approaches the infinite lattice follows from universality, as argued in \cite{ScullardJacobsen2012}, the rapidity of this convergence, especially when compared with standard techniques, is very surprising.

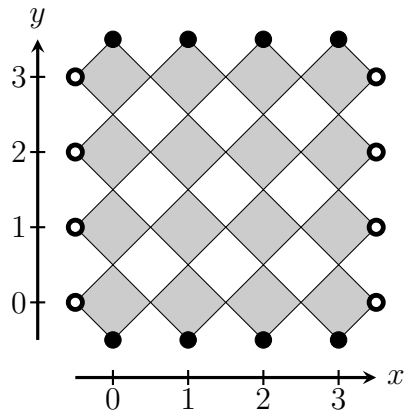
\begin{figure}
\begin{center}

\begin{tikzpicture}[scale=1.0,>=stealth]
\foreach \xpos in {0,1,2,3}
\foreach \ypos in {0,1,2,3}
 \fill[black!20] (\xpos+0.5,\ypos) -- (\xpos+1,\ypos+0.5) -- (\xpos+0.5,\ypos+1) -- (\xpos,\ypos+0.5) -- cycle;
\foreach \xpos in {0,1,2,3}
\foreach \ypos in {0,1,2,3}
 \draw[black] (\xpos+0.5,\ypos) -- (\xpos+1,\ypos+0.5) -- (\xpos+0.5,\ypos+1) -- (\xpos,\ypos+0.5) -- cycle;

\foreach \xpos in {0,1,2,3}
{
 \draw[fill] (\xpos+0.5,0) circle(0.6ex);
 \draw[fill] (\xpos+0.5,4) circle(0.6ex);
}

\foreach \ypos in {0,1,2,3}
{
 \draw[line width=2pt] (0,\ypos+0.5) circle(0.5ex);
 \draw[line width=2pt] (4,\ypos+0.5) circle(0.5ex);
 \draw[fill,white,line width=0pt] (0,\ypos+0.5) circle(0.3ex);
 \draw[fill,white,line width=0pt] (4,\ypos+0.5) circle(0.3ex);
}

\draw[very thick,->] (0,-0.5)--(4,-0.5);
\draw (4,-0.5) node[right] {$x$};
\foreach \xpos in {0,1,2,3}
{
 \draw[thick] (\xpos+0.5,-0.6)--(\xpos+0.5,-0.4);
 \draw (\xpos+0.5,-0.5) node[below] {$\xpos$};
}

\draw[very thick,->] (-0.5,0)--(-0.5,4);
\draw (-0.5,4) node[above] {$y$};
\foreach \ypos in {0,1,2,3}
{
 \draw[thick] (-0.6,\ypos+0.5)--(-0.4,\ypos+0.5);
 \draw (-0.5,\ypos+0.5) node[left] {$\ypos$};
}
 
\end{tikzpicture}
 \caption{Square basis of size $n \times n$ with $n=4$. Open circles indicate the periodic boundary conditions, while the solid circles are terminals. Critical polynomials on all the Archimedean lattices can be computed on these bases, and we use our parallel calculation to find percolation thresholds for $n=8$ and full Potts-model polynomials in $(q,v)$ for $n=6$.}
  \label{fig:square-basis}
\end{center}
\end{figure}

Using a transfer matrix algorithm, one of us \cite{Jacobsen14} computed the critical polynomials in $(q,v)$ of all the Archimedean lattices for square bases of size $5 \times 5$, and computed individual percolation and Potts thresholds for $7 \times 7$. An example of such a basis is shown in Figure \ref{fig:square-basis} and, as explained in \cite{Jacobsen14}, all the Archimedean lattices can be handled with such bases. The accuracy of these estimates is orders of magnitude greater than that obtained with standard techniques, such as Monte Carlo. In this paper, we implement a parallel version of the algorithm in \cite{Jacobsen14} to push the basis size to $n=8$ for percolation thresholds and to $n=6$ for phase diagrams of all the Archimedean lattices. Many interesting features of these phase diagrams become apparent when we examine them for various $n$ and it is surely safe to say that at this point our ability to numerically compute these plots far outstrips our theoretical understanding of them.

\section{Transfer matrix}

The most efficient way to compute the critical polynomial is to use a transfer matrix algorithm \cite{ScullardJacobsen2012,JacobsenScullard2013}. Our earlier transfer matrix computations of critical polynomials would compute the weights of the 0D, 2D and 1D configurations and we could then set $P_{2D}=P_{0D}$. However, this is wasteful because $P_{1D}$ is never used for anything. The trick then would seem to be efficiently generating the weights while discarding the 1D configurations on the fly. This problem is partially solved by Jacobsen's algorithm \cite{Jacobsen14} by recasting the problem in terms of a loop model on square bases that are periodic in the $x$-direction (Figure~\ref{fig:square-basis}). 
At the same time the number of terminals is effectively reduced, leading to further significant gains in performance.

In Figure~\ref{fig:square-basis} each grey square can contain an arbitrary arrangement of vertices and edges, provided that the interaction with the remainder of the basis takes place only through the four vertices at the corners. The transfer matrix contains an operator $\check{\sf R}_i$ materialising the propagation through each grey square at position $x=i$, and $\check{\sf R}_i$ factorises as a product of simpler operators, ${\sf H}_i$ and ${\sf V}_i$, responsible for the addition of one `horizontal' or `vertical' edge to the lattice. Finally, ${\sf H}_i$ and ${\sf V}_i$ are simply expressed in terms of elementary operators ${\sf E}_i$ which are the generators of a periodic Temperley-Lieb algebra \cite{TemperleyLieb71}. The relevant factorised expressions for $\check{\sf R}_i$ corresponding to any Archimedean lattice have been given in \cite{Jacobsen14} and below we shall just use those expressions, with however one significant improvement for the cross lattice (see section~\ref{sec:cross}).

This setup just outlined allows us to discard 1D configurations that wrap through the periodic direction. We still end up computing {\it some} 1D weights but the size of the transfer matrix is greatly reduced over the more na\"ive approach in \cite{ScullardJacobsen2012, JacobsenScullard2013}. Again, we refer the reader to \cite{Jacobsen14} for a detailed description of this improved transfer matrix computation.

\section{Archimedean lattices}
\label{sec:lattices}

\begin{figure}
\begin{center}
 \includegraphics[width=12cm]{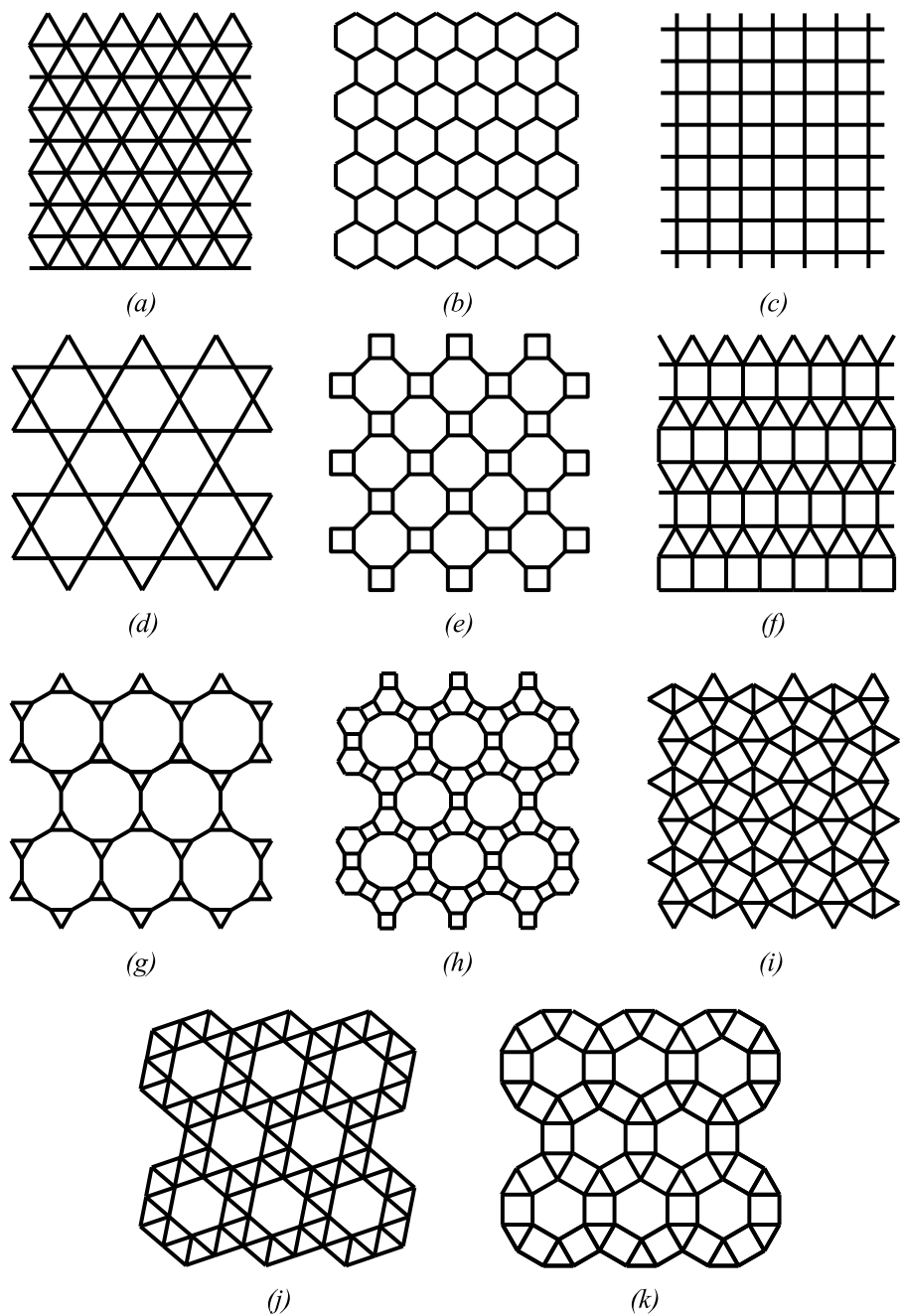}
 \caption{The eleven Archimedean lattices. Their names are given in Table~\ref{tab:archi}.}
 \label{fig:archi}
\end{center}
\end{figure}

The eleven Archimedean lattices are shown in Figure~\ref{fig:archi}. By definition, an Archimedean
lattice is such that each vertex is surrounded by the same types of faces, appearing in the
same cyclic order.

The Potts model is exactly solvable on the square, triangular and hexagonal lattices. We henceforth
focus on the remaining eight cases which are unsolved.

\begin{table}
\begin{center}
 \begin{tabular}{l|ll|lll}
       & Lattice & Notation & Vertices & Edges & Parity of $n$ \\ \hline
 (d) & Kagome & $(3,6,3,6)$ & $3 n^2$ & $6 n^2$ & Any \\
 (e) & Four-eight & $(4,8^2)$ & $4 n^2$ & $6 n^2$ & Any \\
 (f) & Frieze & $(3^3,4^2)$ & $2 n^2$ & $5 n^2$ & Any \\
 (g) & Three-twelve & $(3,12^2)$ & $6 n^2$ & $9 n^2$ & Any \\
 (h) & Cross & $(4,6,12)$ & $6 n^2$ & $9 n^2$ & Even \\
 (i) & Snub square & $(3^2,4,3,4)$ & $2 n^2$ & $5 n^2$ & Even \\
 (j) & Snub hexagonal & $(3^4,6)$ & $3 n^3$ & $\frac{15}{2} n^2$ & Even \\
 (k) & Ruby & $(3,4,6,4)$ & $3 n^2$ & $6 n^2$ & Even \\
 \end{tabular}
 \caption{Nomenclature of the Archimedean lattices and their duals (Laves lattices).
 The labels (d)--(k) refer to the eight lattices (see Figure~\ref{fig:archi}) on which
 the $q$-state Potts model is unsolved.
 The notation is that of Gr\"unbaum and Shephard \cite{GrunbaumShephard87}.
 The right part of the table gives the number of vertices and edges for square
 bases of size $n \times n$ grey squares. In addition we state any parity constraints on $n$.
 Note that the cross lattice comprises an improvement over \cite[Table 3]{Jacobsen14}.}
 \label{tab:archi}
\end{center}
\end{table}

\section{Parallelisation}

The algorithm used in this work is a parallelised version of that described in detail in \cite{Jacobsen14}. At its heart, the computation is the multiplication of a vector by a matrix, and the parallel strategy is simply to distribute the vector among the processors such that each contains an equal number of states. Then if the result of an elementary ${\sf E}_i$ operation produces a state that belongs on a different processor, that state's weight is sent to its appropriate destination. Unfortunately, there is no way to partition the states such that no communication is needed between the processors, so the goal of a good parallelisation scheme is to minimise the amount of inter-processor communication.

To acheive this goal we modify slightly the ordering of the states from the one given in \cite{Jacobsen14}. Here we take care to ensure that when implementing the bijection between connectivity states and integers, the reduced state describing the bottom row is the most slowly varying.%
\footnote{This is done simply by changing the order of priorities between points (ii) and (iv) appearing in the itemised list in \cite[section~3.5.4]{Jacobsen14}.}
This ensures that blocks of states with the same bottom-row connectivity will tend to be placed on the same processor. Because the action of the transfer matrix is implemented with ${\sf H}_i$ and ${\sf V}_i$ operators acting on the top row, the bottom row is often unchanged upon perfoming these operations. Even so, the connectivity on the bottom row might still change if two strings are attached,%
\footnote{See point (vi) in the second itemised list of \cite[section~3.5.2]{Jacobsen14}.}
but this is rare enough that ordering the states in the way we describe significantly cuts down on the inter-processor communication. Beyond this, we take no other care in distributing states across tasks.
If the system has $N$ total states, and we use $p$ processors, $p$ is chosen as a divisor of $N$ and the first processor gets the first $N/p$ states, the second the next $N/p$, etc.

In the serial code, $n=5$ was the maximum size accessible for computing full Potts-model polynomials in $(q,v)$. In \cite{Jacobsen14}, the transfer matrix was also used to compute individual thresholds for $q=1,2$ and $3$ using a Newton-Raphson technique and $n=7$ was the limit for computing these. Using the parallel algorithm we push these limits to $n=6$ for the full polynomials and $n=8$ for $q=1$ (percolation) critical points, giving us the clearest look yet at the phase diagrams of all the Archimedean lattices and earning typically an extra digit of accuracy in the percolation thresholds.

To compute the full polynomials, whose coefficients are very large positive integers, we perform the calculation modulo a set of prime numbers. The results for each prime are then combined into the final answer using the Chinese remainder theorem. However, to compute just the percolation thresholds for $n=8$, we use the real numbers provided by the CLN arbitrary precision library for C++ \cite{CLN}. CLN objects are equipped with functions for writing numbers to and from strings, and to communicate between processors we send the string and convert it to a CLN number at the destination. This adds somewhat to the usual communication cost.

Although the transfer matrix has the same dimension for a given $n$, independent of the lattice, there is significant variation in the numbers of edges and vertices in each problem. Therefore, computing the full $(q,v)$ polynomial requires more resources, in terms of both time and memory, for some Archimedean lattices than others, with the $(3,12^2)$ and cross lattices being the most difficult. The number of processors we chose for each problem is governed both by memory and speed considerations. For the $n=8$ percolation thresholds we used 4004 processors in all cases, with the time the runs took depending on the lattice and the quality of the initial guess.

We performed these runs on the Cab and Vulcan supercomputers at Lawrence Livermore National Laboratory. 

\section{Percolation thresholds on Archimedean lattices}
\label{sec:pc-archi}

In Ref.~\cite{Jacobsen14} it was shown how each of the Archimedean lattices can be represented in the
form shown in Figure \ref{fig:square-basis}, enabling a transfer matrix calculation of the graph polynomial $P_B(q,v)$
by specifying a corresponding $\check{\sf R}$-matrix. We use here the same bases, except for the cross
lattice for which we have now found a better representation containing twice as many vertices and edges
for any given $n$.

For $q=1$, the root $v_{\rm c}$ of $P_B(1,v)$ that approximates the percolation threshold $p_{\rm c}$,
via the relation $p = v/(1+v)$, was obtained in Ref.~\cite{Jacobsen14} for sizes $n=1,2,\ldots,7$.
For each of the eight unsolved Archimedean lattices we have extended this computation to $n=8$;
see Table~\ref{tab:pc}.

\begin{table}
\begin{center}
 \begin{tabular}{l|l}
 Lattice & $p_{\rm c}$ \\ \hline
 Kagome             & 0.52440499951414108517718152880414539618186347478997 \\
 Four-eight          & 0.67680315546810921539888642691310355045728163134828 \\
 Frieze                & 0.41964027675236652348940526360240730997388408890937 \\
 Three-twelve      & 0.74042079889005115125415547646085000986933464791292 \\
 Cross                 & 0.69373316681403389479582584979836234112904294069266 \\ 
 Snub square      & 0.41413788425992466879112356169148391265046003220755 \\
 Snub hexagonal & 0.43432836999043911236626317783310197323151394980874 \\
 Ruby                  & 0.52483149832115089795511006350993499311132979786833 \\
 \end{tabular}
 \caption{Bond percolation thresholds $p_{\rm c}(n)$ for $n \times n$ bases of size $n=8$ on the non-solvable Archimedean lattices.}
  \label{tab:pc}
\end{center}
\end{table}

For each lattice the behaviour of $p_{\rm c}(n)$ appears to have the form
\begin{equation}
 p_{\rm c}(n) = p_{\rm c} + A n^{-w} \,.
\label{extrapol_w}
\end{equation}
Approximate values of the exponent $w$ can be obtained from non-linear three-point fits to the data for
each lattice. These estimates for $w$ are shown in Table~\ref{tab:w_pc}.
The most reliable values are obtained for the lattice for which there is no parity
constraint on $n$ (see Table~\ref{tab:archi}), and in those cases we provide an error bar
in Table~\ref{tab:w_pc} which is obtained by comparing the three-point fits for $n \in \{6,7,8\}$
and for $n \in \{5,6,7\}$. In the cases with parity constraint, we just give the value for the fit
with $n \in \{4,6,8\}$ to one decimal place.

\begin{table}
\begin{center}
 \begin{tabular}{l|ll}
 Lattice & $w$ & $p_{\rm c}$ \\ \hline
 Kagome             & 6.346 (2) & 0.52440499917 (1) \\
 Four-eight          & 4.20 (8) & 0.676803125 (2)  \\
 Frieze                & 3.5 & 0.41964037 (2) \\
 Three-twelve      & 6.392 (1) & 0.7404207988509 (4) \\
 Cross                 & 4.7 & 0.69373295 (9) \\ 
 Snub square      & 4.1 & 0.414137858 (1) \\
 Snub hexagonal & 5.1 & 0.43432830 (2) \\
 Ruby                  & 4.9 & 0.52483144 (2) \\
 \end{tabular}
 \caption{Estimates for the finite-size scaling exponent $w$ of (\ref{extrapol_w}), and final results for the percolation thresholds $p_{\rm c}$, for bond percolation on the non-solvable Archimedean lattices.}
 \label{tab:w_pc}
\end{center}
\end{table}

It appears that the lattices with three-fold rotational symmetry (kagome, three-twelve, cross,
snub hexagonal, and ruby) might have a common value of $w$ that we can estimate as
\begin{equation}
 w_3 = 6.35 \pm 0.05 \,.
 \label{winfty3}
\end{equation}
independently of the lattice. The remaining lattices which have four-fold symmetry
(four-eight, snub square) or two-fold symmetry (frieze) appear to share another value of
$w$ that we estimate as
\begin{equation}
 w_4 = 4.2 \pm 0.1 \,.
 \label{winfty4}
\end{equation}

To extrapolate $p_{\rm c}(n)$ to the thermodynamic limit, $n \to \infty$, we use Monroe's
implementation \cite{Monroe} of the Bulirsch-Stoer (BS) extrapolation scheme \cite{BulirschStoer}.
This assumes that the data has the form (\ref{extrapol_w}) and we must provide the corresponding
value of $w$. It goes without saying that this procedure would gain substantially 
in precision and reliability if there was an analytic prediction for $w$. In the absense of any
such result, we perform for each lattice the extrapolations using both the $w$ of Table~\ref{tab:w_pc}
and the estimates $w_3$ or $w_4$ (depending on the symmetry class). The final results and
error bars on $p_{\rm c}$, obtained by comparing the last few BS extrapolants for both values
of $w$, are displayed in the right column of Table~\ref{tab:w_pc}.

We should stress that this extrapolation procedure is somewhat more cautious than the one used
in Ref.~\cite{Jacobsen14} and accordingly produces more conservative error bars for series of
the same length. The fact that we have now one more data point for each lattice obviously
makes the thresholds of Table~\ref{tab:w_pc} more reliable and/or more precise than those
in Ref.~\cite{Jacobsen14}.

\section{Phase diagrams}
\label{sec:pd}

The full polynomial $P_B(q,v)$ was obtained in symbolic form for $n=1,2,\ldots,5$ in Ref.~\cite{Jacobsen14}. We have extended these computations to $n=6$. As in our preceding work \cite{ScullardJacobsen2012,JacobsenScullard2013,Jacobsen14} the polynomials are available in electronic form
as supplementary material to this paper.%
\footnote{The text file {\tt PB6.m} provided can be processed by {\sc Mathematica} or, perhaps after minor changes of formatting, by
any symbolic computer algebra program of the reader's liking.}

The zero set $P_B(q,v) = 0$ in the real $(q,v)$ plane was shown previously
\cite{Jacobsen12,ScullardJacobsen2012,JacobsenScullard2013,Jacobsen14} to provide precise information about the phase diagram of the corresponding Potts models. In the next sections we show these diagrams for the unsolved Archimedean lattices.

Actually, even for solved problems, the set of solvable curves does not give access to all relevant information in
the antiferromagnetic region $v<0$. It was argued in \cite{Jacobsen12} that the roots of
the critical polynomial will produce extra curves inside the so-called Berker-Kadanoff phase \cite{Saleur91}
which will form, in the thermodynamic limit, vertical rays in the $(q,v)$ plane at the Beraha numbers
\begin{equation}
 q = B_k \equiv 4 \cos^2 \frac{\pi}{k} .
\end{equation}
for certain integer $k$. On the basis of conformal field theory results and cognate studies of partition function zeros
\cite{Salas06,Salas07}, as well as actual evidence from critical polynomials, it was conjectured in \cite{JacobsenScullard2013}
that such vertical rays will exist for any even $k=4,6,8,\cdots$, as far as allowed by the extent of the Berker-Kadanoff phase.

For the case of the square lattice the Berker-Kadanoff phase extends all the way to $(q,v) = (4,-2)$,
and explicit calculations of $P_B(q,v)$ with $n$ up to $5$ \cite{Jacobsen14} fully confirmed the validity of the conjecture.
Accordingly, we see no reason to  compute this polynomial also for $n=6$.
However, vertical rays in a Berker-Kadanoff-like phase appear to be a generic feature, present in all the Archimedean lattices,
and the number and position of such rays permit us to shed light on the extent of the Berker-Kadanoff phase for each lattice.
The $n=6$ critical polynomials discussed below will extend the results of \cite{Jacobsen14} in this respect.

We should stress that the physics of the antiferromagnetic regime is very difficult to access using standard numerical techniques,
in particular for the non-probabilistic regime $v < -1$, but critical polynomials capture this regime very well. 

\subsection{Kagome lattice $(3,6,3,6)$}

\begin{figure}
\begin{center}
\includegraphics[width=12cm]{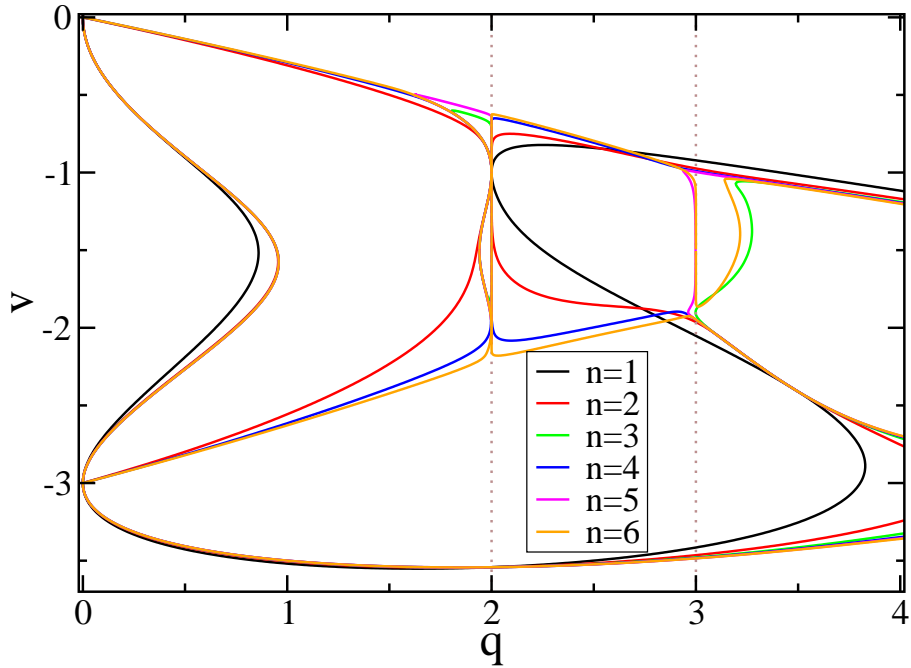}
\caption{Roots of $P_B(q,v)$ for the Potts model on the kagome lattice, using
$n \times n$ square bases. Here and in the following figures, the grey dotted lines indicate the Beraha numbers
for which vertical rays are observed.}
\label{fig:kagome-pd}
\end{center}
\end{figure}

To get the $n=6$ critical polynomial, we needed to perform the calculation with 13 different primes, with each prime taking about 5 hours on 3861 processors. The phase diagram for this lattice is shown in Figure~\ref{fig:kagome-pd} along with those from the smaller bases already reported in \cite{Jacobsen14}. The $n=6$ curve is very similar to the union of the $n=3,4$ and $5$ curves, probably indicating that our $n=6$ result is indeed an accurate picture of the true phase diagram. The only vertical rays to appear are at $q=2$ and $3$, and the $n=6$ polynomial does not give a hint that any further rays are going to appear. Also, the Berker-Kadanoff phase appears as though it may not extend to $(q,v)=(4,-2)$ as it does for most Archimedean lattices. A curious persistent feature is the small bump that rises to the left of the $q=2$ ray. As for most of the features in this and the subsequent phase diagrams, there is not yet any analytical explanation for this.

A close scrutiny of Figure~\ref{fig:kagome-pd} reveals a number of parity effects, confirming and extending observations made in \cite{Jacobsen12,JacobsenScullard2013,Jacobsen14}. For instance, the lower boundary of the Berker-Kadanoff phase---the curve going
emanating from $(q,v) = (0,-3)$ and going towards $\approx (3,-2)$---appears only for even $n$, and a small part of the upper
boundary is visible only for odd $n$. Maybe the most significant feature of the $n=6$ plot is that it confirms the existence of a bubble
to the right of $q=3$, previously seen only for $n=3$. We conjecture that it will appear when $n$ is any multiple of $3$.

To get the $n=8$ percolation estimate took about 6.5 hours on 4004 processors.

\subsection{Four-eight lattice $(4,8^2)$}

\begin{figure}
\begin{center}
\includegraphics[width=12cm]{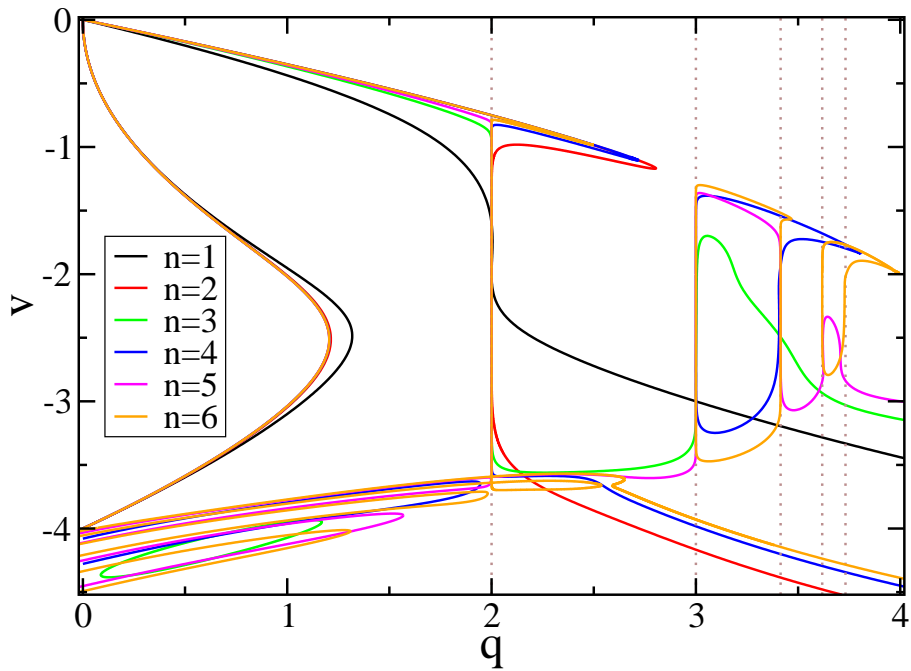}
\caption{Roots of $P_B(q,v)$ for the Potts model on the four-eight lattice, using
$n \times n$ square bases.}
\label{fig:foureight-pd}
\end{center}
\end{figure}

Although the unit cell of the $(4,8^2)$ lattice has the same number of edges as the kagome lattice, it has more vertices and therefore the $n=6$ polynomial is slightly more difficult to compute due to its higher order in $q$ (144 vs.\ 108). To do the computation with one prime took around 7 hours on 4092 processors and we used $14$ primes. The phase diagram arising from the $n=6$ polynomial is shown in Figure \ref{fig:foureight-pd} along with those for $n$ up to $5$ reported previously in \cite{Jacobsen14}. Unlike in the kagome case, we have vertical rays here at several Beraha numbers ($B_4, B_6,\ldots,B_{12}$) and it certainly seems plausible that more would appear if we could compute higher order polynomials. In fact, this phase diagram is very similar to that of the square lattice, which is perhaps not surprising. Here, though, we have some additional features such as the finger-like structures that emanate from the lower left of the diagram. The line that exits through the lower right (for even $n$ only) is similarly not present in the square lattice.

\subsection{Frieze lattice $(3^3,4^2)$}

\begin{figure}
\begin{center}
\includegraphics[width=12cm]{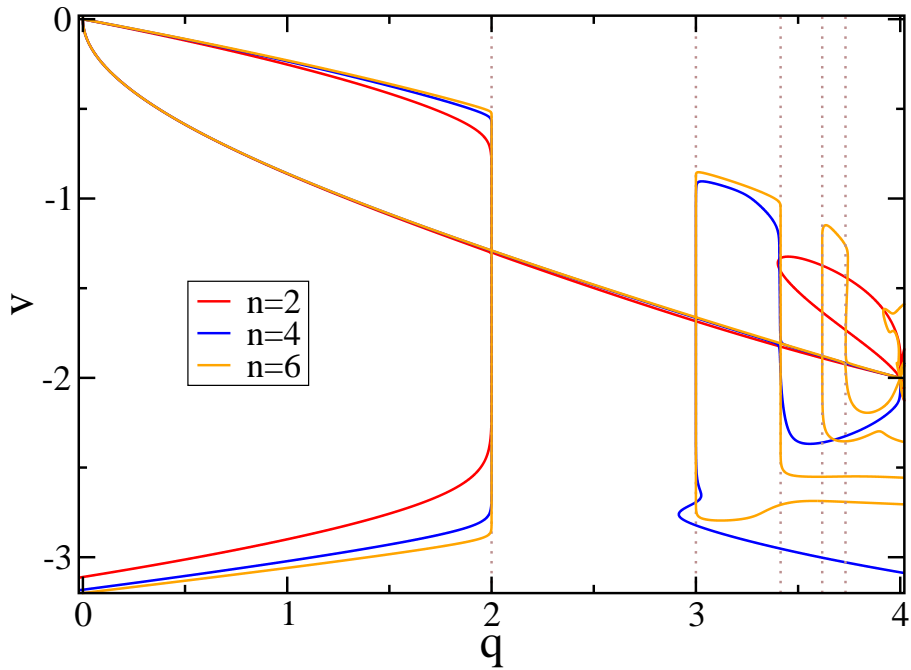}
\caption{Roots of $P_B(q,v)$ for the Potts model on the frieze lattice, using
$n \times n$ square bases.}
\label{fig:frieze-pd}
\end{center}
\end{figure}

For the frieze lattice we used 4092 processors, and completed a computation for a single prime in about 3 hours. We used 14 primes, but 13 were sufficient to get the final answer. For this problem, the lattice only fits into our basis for even $n$. The $n=6$ phase diagram is plotted in Figure \ref{fig:frieze-pd} along with the $n=2$ and $4$ results from \cite{Jacobsen14}. In this and subsequent cases, where only even $n$ are possible, the $n=6$ polyomial adds significant detail to the phase diagram. In particular we observe two more vertical rays at the Beraha numbers $B_{10}$ and $B_{12}$. Note that there seem to be gaps in the boundary of the Berker-Kadanoff phase for all $n$ between $q=2$ and $3$. These are probably not a real feature but an effect of the basis we are using. Doing the calculation for a different basis, such as one in which we shift the identifications between the top and bottom rows, may fill in these gaps.

\subsection{Three-twelve lattice $(3,12^2)$}

\begin{figure}
\begin{center}
\includegraphics[width=12cm]{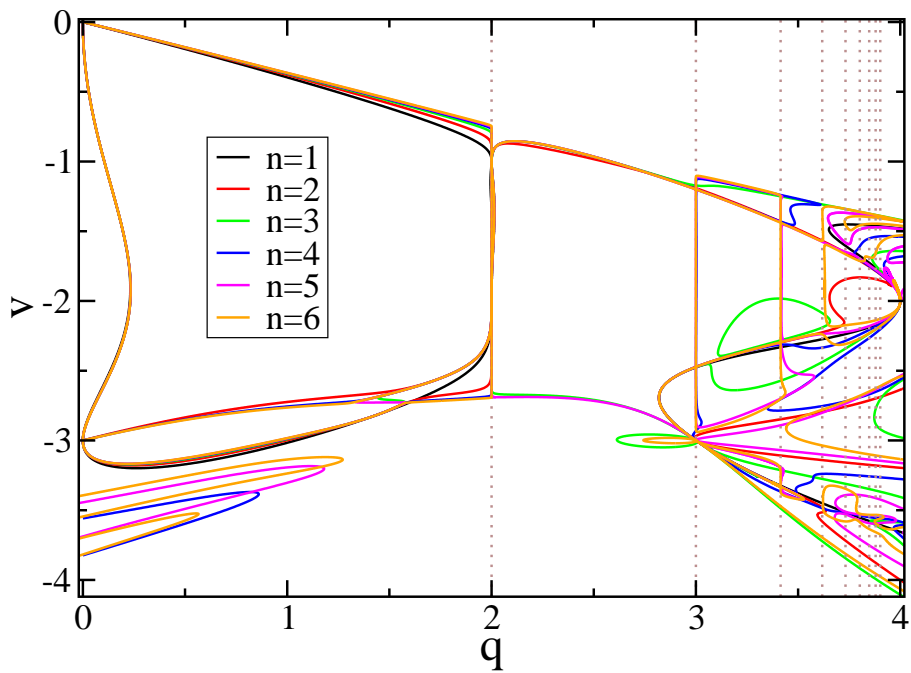}
\caption{Roots of $P_B(q,v)$ for the Potts model on the three-twelve lattice, using
$n \times n$ square bases.}
\label{fig:threetwelve-pd}
\end{center}
\end{figure}

The $n=6$ polynomial for the $(3,12^2)$ lattice is order $324$ in $v$ and $216$ in $q$. This makes the computation difficult both in terms of the number of primes needed, 22, and the time it takes to do a calculation for a single prime, 9 hours. However, we are rewarded for this calculation with the very intricate phase diagram shown in Figure \ref{fig:threetwelve-pd} (along with those from previous work \cite{Jacobsen14,JacobsenScullard2013}).

Although this lattice bears some similarity to the kagome lattice, aside from the small bump (difficult to see at the scale of the plot) rising off the $q=2$ ray the phase diagram is quite different. The Berker-Kadanoff phase clearly extends to $(q,v)=(4,-2)$ and has vertical rays at at least four Beraha numbers ($B_4, B_6, B_8, B_{10}$). But a closer scrutiny of Figure~\ref{fig:threetwelve-pd} reveals emergent features at five more Beraha numbers ($B_{12}, B_{14},\ldots,B_{20}$) that would presumably turn into fully fledged rays if further sizes $n>6$ were available.
We also have the fingers that emerge from the lower left, while the region outside the Berker-Kadanoff phase for $q>3$ seems to be a rather wild tangle.

\subsection{Cross lattice $(4,6,12)$}
\label{sec:cross}

\begin{figure}
\begin{center}
\begin{tikzpicture}[scale=1.0,>=stealth]
\foreach \xpos in {0,1,2,3}
\foreach \ypos in {0,2,4,6}
 \fill[black!20] (\xpos+0.5,\ypos) -- (\xpos+1,\ypos+0.5) -- (\xpos+0.5,\ypos+1) -- (\xpos,\ypos+0.5) -- cycle;

\foreach \xpos in {0,2}
\foreach \ypos in {0,4}
{
 \fill[red!20] (\xpos+0.5,\ypos+1) -- (\xpos+1,\ypos+0.5) -- (\xpos+1.5,\ypos+1) -- (\xpos+1.5,\ypos+2) -- (\xpos+1,\ypos+2.5) -- (\xpos+0.5,\ypos+2) -- cycle;
 \fill[red!20] (\xpos,\ypos+2.5) -- (\xpos+0.5,\ypos+3) -- (\xpos+0.5,\ypos+4) -- (\xpos,\ypos+4.5) -- cycle;
 \fill[red!20] (\xpos+2,\ypos+2.5) -- (\xpos+1.5,\ypos+3) -- (\xpos+1.5,\ypos+4) -- (\xpos+2,\ypos+4.5) -- cycle;
}

\foreach \xpos in {0,1,2,3}
\foreach \ypos in {0,2,4,6}
{
  \draw[blue,ultra thick] (\xpos,\ypos+0.5) -- (\xpos+0.5,\ypos) -- (\xpos+0.5,\ypos+1) -- cycle;
  \draw[blue,ultra thick] (\xpos+0.5,\ypos+0.5) -- (\xpos+1,\ypos+0.5);
  \draw[blue,ultra thick] (\xpos+0.5,\ypos+1) -- (\xpos+0.5,\ypos+2);
}

\foreach \xpos in {0,2}
\foreach \ypos in {0,4}
{
 \draw[blue,ultra thick] (\xpos+0.5,\ypos+1.33) -- (\xpos+1.5,\ypos+1.33);
 \draw[blue,ultra thick] (\xpos+0.5,\ypos+1.67) -- (\xpos+1.5,\ypos+1.67);
 \draw[blue,ultra thick] (\xpos+0.5,\ypos+3.33) -- (\xpos,\ypos+3.33);
 \draw[blue,ultra thick] (\xpos+0.5,\ypos+3.67) -- (\xpos,\ypos+3.67);
 \draw[blue,ultra thick] (\xpos+1.5,\ypos+3.33) -- (\xpos+2,\ypos+3.33);
 \draw[blue,ultra thick] (\xpos+1.5,\ypos+3.67) -- (\xpos+2,\ypos+3.67);
}


\draw[very thick,->] (0,-0.5)--(4,-0.5);
\draw (4,-0.5) node[right] {$x$};
\foreach \xpos in {0,1,2,3}
{
 \draw[thick] (\xpos+0.5,-0.6)--(\xpos+0.5,-0.4);
 \draw (\xpos+0.5,-0.5) node[below] {$\xpos$};
}

\draw[very thick,->] (-0.5,0)--(-0.5,8);
\draw (-0.5,8) node[above] {$y$};
\foreach \ypos in {0,1,2,3}
{
 \draw[thick] (-0.6,2*\ypos+0.5)--(-0.4,2*\ypos+0.5);
 \draw (-0.5,2*\ypos+0.5) node[left] {$\ypos$};
}
 
\end{tikzpicture}
 \caption{Four-terminal representation of the cross lattice.}
 \label{fig:cross}
\end{center}
\end{figure}
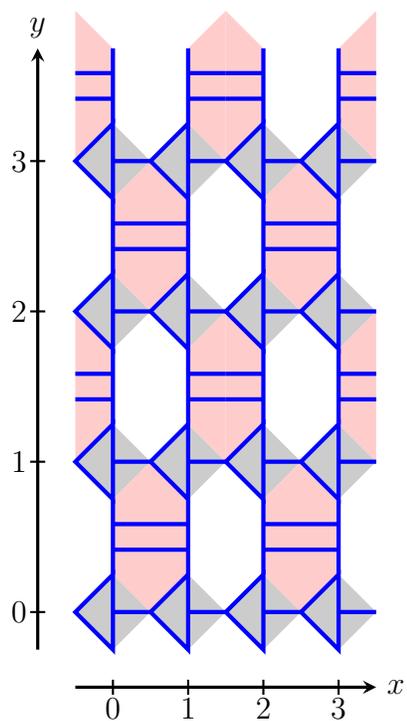

For the cross lattice we have found a four-terminal representation which for a given $n$
contains twice the number of vertices and edges than the representation used in
Ref.~\cite{Jacobsen14}. This is accomplished by choosing the following $\check{\sf R}$-matrix:
\begin{equation}
 \check{\sf R}_i = {\sf V}_{i+2} {\sf H}_{i+1} {\sf V}_i {\sf V}_{i+2} {\sf H}_{i+1} \,,
 \label{cross-R}
\end{equation}
whose action is illustrated in the grey squares of Figure~\ref{fig:cross}.
In addition we use the trick that was dubbed ``white hexagons'' in \cite[section 4.9]{Jacobsen14}
to insert extra structure in-between two rows of the basis. The corresponding operator reads
\begin{equation}
 {\sf O}_i = ({\sf V}_i {\sf V}_{i+2} {\sf H}_{i+1})^2  {\sf V}_i {\sf V}_{i+2}
\end{equation}
and its action is illustrated in the pink hexagons (dubbed ``white'' in \cite{Jacobsen14}) of Figure~\ref{fig:cross}.

\begin{figure}
\begin{center}
\includegraphics[width=12cm]{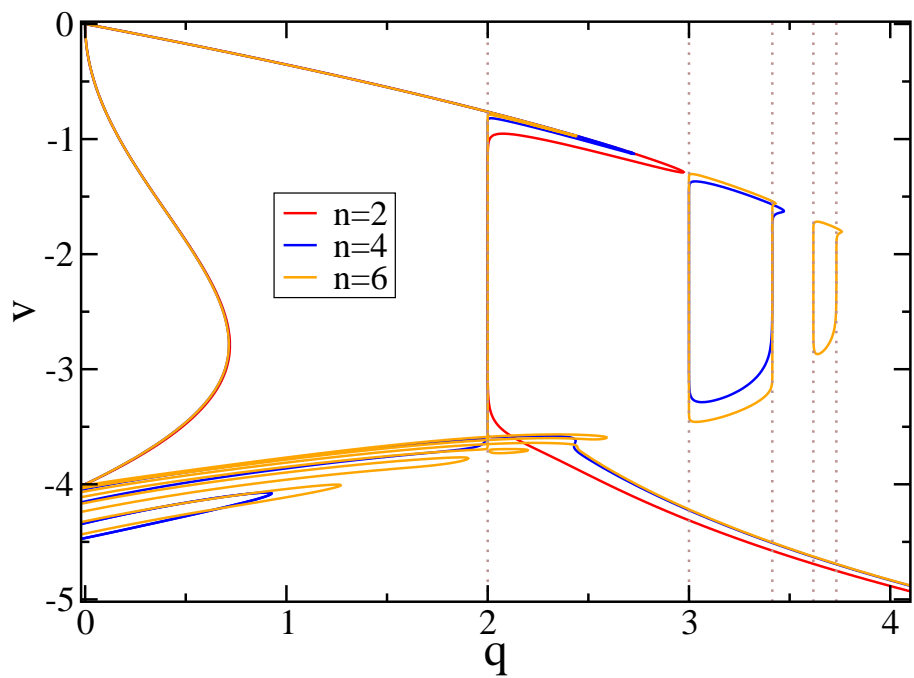}
\caption{Roots of $P_B(q,v)$ for the Potts model on the cross lattice, using
$n \times n$ square bases.}
\label{fig:cross-pd}
\end{center}
\end{figure}

Computing this polynomial requires a similar effort as that of the $(3,12^2)$ lattice; and like the frieze lattice, the cross lattice only fits into bases with even $n$. The phase diagram is shown in Figure \ref{fig:cross-pd}; the curves with $n=2$ and $n=4$ this time do {\em not} correspond to those computed in \cite{Jacobsen14} but represent the new construction (\ref{cross-R}) applied to those two cases. We can see vertical rays at the Beraha numbers, with $n=6$ providing two more ($B_{10}$ and $B_{12}$) above $n=4$. We also have the fingers that enter through the lower left and the line that exits through the lower right. Overall, the phase diagram is qualitatively very similar to that of the $(4,8^2)$ lattice.

\subsection{Snub square lattice $(3^2,4,3,4)$}
\begin{figure}
\begin{center}
\includegraphics[width=12cm]{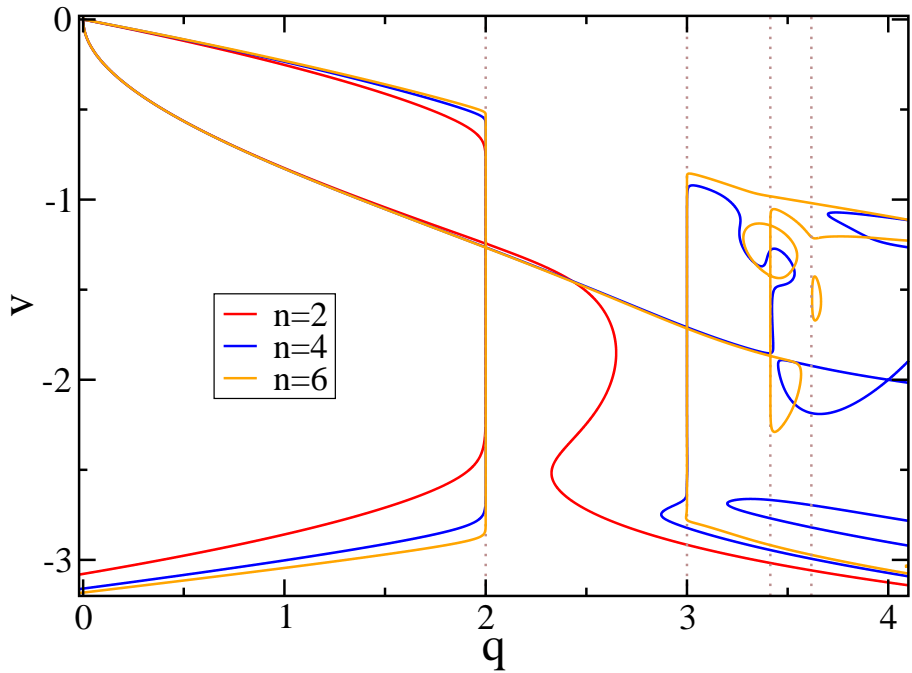}
\caption{Roots of $P_B(q,v)$ for the Potts model on the snub square lattice, using
$n \times n$ square bases.}
\label{fig:snubsquare-pd}
\end{center}
\end{figure}

Calculating the $n=6$ critical polynomial for the snub square lattice required 11 primes. We used 2046 processors with which it took about 3 hours for each prime. The phase diagram is shown in Figure \ref{fig:snubsquare-pd}. Along with the usual vertical rays at the even Beraha numbers, there is an oval-shaped region straddling the one at $B_8$ near $v \approx -1.5$. It appears that the Berker-Kadanoff may not extend to $(q,v)=(4,-2)$.

\subsection{Snub hexagonal lattice $(3^4,6)$}
\begin{figure}
\begin{center}
\includegraphics[width=12cm]{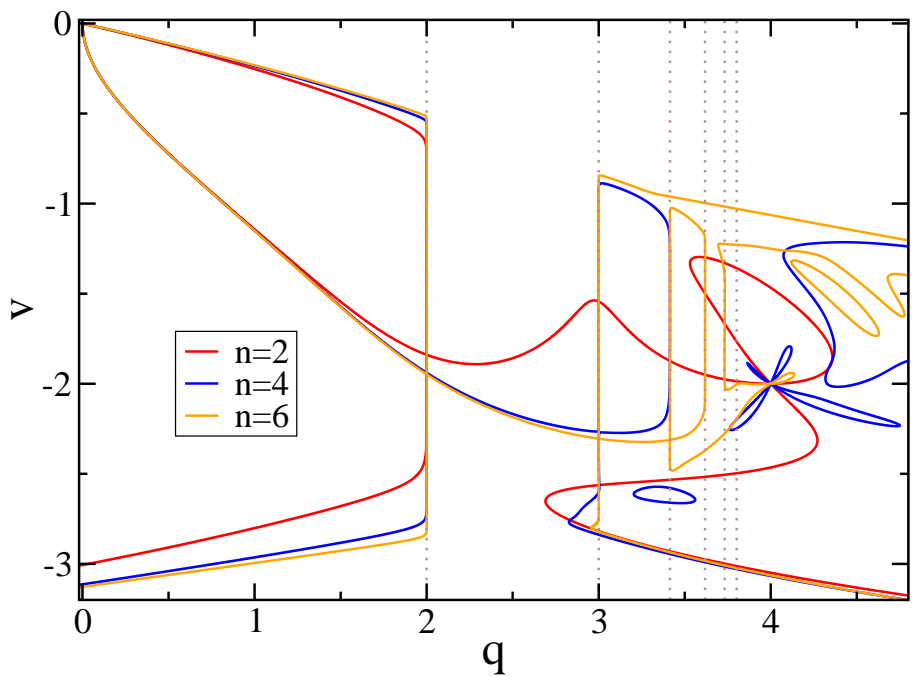}
\caption{Roots of $P_B(q,v)$ for the Potts model on the snub hexagonal lattice, using
$n \times n$ square bases.}
\label{fig:snubhexagonal-pd}
\end{center}
\end{figure}

To do the calculation of the snub hexagonal polynomial requires about 7 hours on 4092 processors. We used 17 primes. The phase diagram is shown in Figure \ref{fig:snubhexagonal-pd}. As for other lattices, we can only find the polynomial for even $n$, and there is a gap in the boundary of the Berker-Kadanoff phase between $q=2$ and $3$. This phase appears to reach $(q,v)=(4,-2)$ where there is an interesting flower-like structure in the $n=4$ and $6$ polynomials.

\begin{figure}
\begin{center}
\includegraphics[width=12cm]{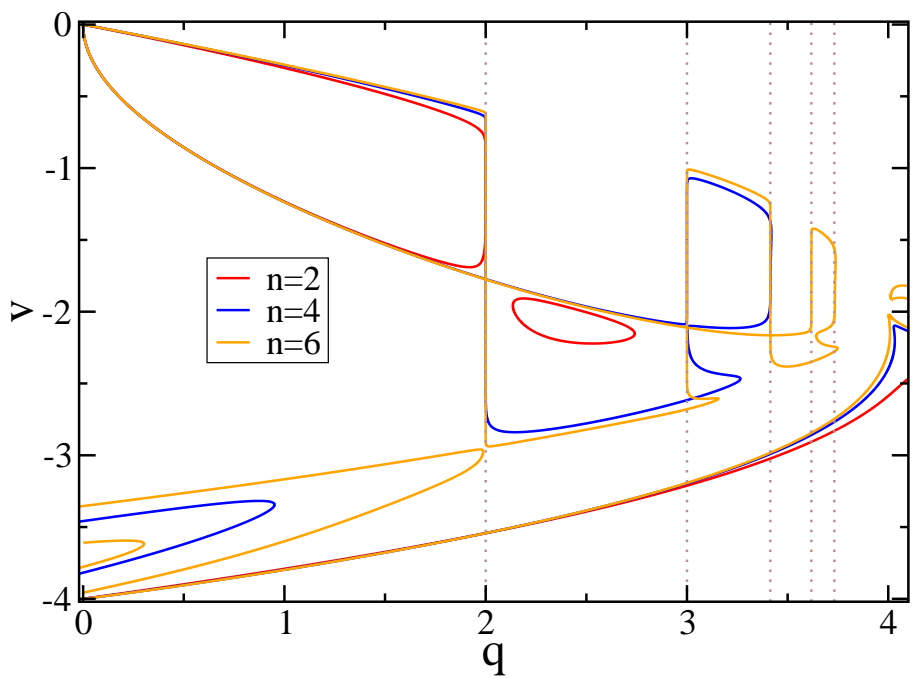}
\caption{Roots of $P_B(q,v)$ for the Potts model on the ruby lattice, using
$n \times n$ square bases.}
\label{fig:ruby-pd}
\end{center}
\end{figure}

\subsection{Ruby lattice $(3,4,6,4)$}
For the ruby lattice, we used 2046 processors; it took about 5 hours to do the computation for a single prime, and we used 14 primes. Once again, only even $n$ works for this problem. The phase diagrams are shown in Figure \ref{fig:ruby-pd}. Our $n=6$ calculation adds significant detail, such as the two rays at $B_{10}$ and $B_{12}$. The Berker-Kadanoff phase likely extends to $(q,v)=(4,-2)$.

\section{Discussion}
\label{sec:disc}

We have implemented a parallel version of the improved transfer matrix algorithm given in \cite{Jacobsen14} for computing critical polynomials. We have computed phase diagrams for bases of size $6 \times 6$ and found percolation thresholds for $8 \times 8$ bases. Using Bulirsch-Stoer extrapolation we have found the latter to accuracies in some cases far exceeding that possible with standard techniques; see Table~\ref{tab:w_pc}. Even with large parallel calculations, as we have done, Monte Carlo and similar techniques would obtain nowhere near the accuracy found here for our best results (references to other work were given in the tables of Ref.~\cite{Jacobsen14}). On the other hand, there are still many non-Archimedean two-dimensional problems for which it is not yet clear how to better Monte Carlo with critical polynomials (see, e.g., \cite{Damavandi}).

For the phase diagrams, we are given a view of the $v<-1$ regime of the Potts model that is accessible only with critical polynomials, and in this work we have the most detailed picture yet for the Archimedean lattices. However, only the square lattice phase diagram is properly understood analytically \cite{Jacobsen14,Jacobsen12,Saleur91}, and, apart from some qualitative similarities with that case, the many interesting features in these plots are yet to be explained.

Aside from these issues, many interesting theoretical questions remain to be addressed about the method itself. For example, it would be very useful to know the exponent $w$ in the scaling law (\ref{extrapol_w}) exactly. It seems clear that, unlike most other exponents that arise in these models, $w$ is not universal; it takes a value around $6$ for some lattices and $4$ for others, possibly depending on the degree of symmetry. Gaining some understanding of this exponent may help to answer the broader question of why this method performs as well as it does.

\section*{Acknowledgments}
CRS acknowledges the helpful advice of Jim Glosli at LLNL in programming the parallel implementation. This work was partially performed under the auspices of the U.S. Department of Energy at the Lawrence Livermore National Laboratory under Contract No. DE-AC52-07NA27344. The research of JLJ is supported by the Agence Nationale de la Recherche (grant ANR-10-BLAN-0414: DIME) and the Institut Universitaire de France.

\subsection*{Note added}

After the completion of this work, one of us conceived an alternative eigenvalue method
for computing the critical polynomial on semi-infinite bases \cite{Jacobsen15}. This method
sheds more light on the scaling exponents $w$ and leads to further gain in precision. For
instance, our value and error bar of the bond percolation threshold on the kagome lattice
(see Table~\ref{tab:w_pc}) are confirmed by the new method which gives
$p_{\rm c} = 0.524\, 404\, 999\, 167\, 439(4)$. Work on parallelising the eigenvalue method
is forthcoming.

\section*{References}
\bibliographystyle{iopart-num}
\bibliography{SJ}

\end{document}